\documentclass[aps,preprint,pre,superscriptaddress]{revtex4}

\usepackage{graphicx,color}
\usepackage{ae,aecompl}
\usepackage{amsmath,amssymb,amstext}
\usepackage[latin1]{inputenc}
\usepackage{verbatim}

\begin{document}

\title{Computational Analysis of Composition-Structure-Property-Relationships in NZP-type Materials for Li-Ion Batteries}

\author{Daniel Mutter}
\author{Daniel F. Urban}
\author{Christian Els\"{a}sser}
\affiliation{Fraunhofer Institute for Mechanics of Materials IWM, W\"{o}hlerstra\ss{}e 11, 79108 Freiburg, Germany}
\affiliation{Freiburg Materials Research Center (FMF), Albert-Ludwigs-Universit\"{a}t Freiburg, Stefan-Meier-Stra\ss{}e 21, 79104 Freiburg, Germany}

\begin{abstract}
Compounds crystallizing in the structure of NaZr$_2$(PO$_4$)$_3$ (NZP) are considered as promising materials for solid state electrolytes in Li-ion batteries. Using density functional theory (DFT), a systematic computational screening of 18 NZP compounds, namely LiX$_2$(LO$_4$)$_3$ with X = Ti, V, Fe, Zr, Nb, Ru, Hf, Ta, Os, and L = P, Mn is performed with respect to their activation energies for vacancy-mediated Li migration. It is shown how the different ionic radii of the cationic substitutions influence structural characteristics such as the octahedron volumes around Li ions on the initial and transition state sites, which affect the activation energies (''composition-structure-property'' relationships). The prevalent assumption that structural bottlenecks formed by triangularly arranged oxygen atoms at a certain location along the migration path determine the energy barriers for Li migration is not supported by the DFT results. Instead, the ionic neighborhood of the migrating ion in the initial and in the transition state needs to be taken into account to relate the structure to the activation energies. This conclusion applies to Na containing NZP compounds as well.
\end{abstract}

\maketitle

\section{Introduction}\label{intro}

A major challenge in the development of future portable energy storage devices with the potential to outreach today's battery technology in terms of capacity, energy density, charge rate, cyclability, and safety, is the identification of new materials for the use as anodes, cathodes, or electrolytes with improved properties such as ionic conductivity or thermal, chemical, electrochemical, and mechanical stability \cite{Scrosati2010}. A popular approach in the search for new materials taking advantage of today's available computational resources is the virtual screening and characterization of hundreds of compounds which are systematically generated by substituting elements at specific lattice sites of a material class, with the goal of identifying compositions leading to improved battery-relevant properties \cite{Kirklin2013, Ling2016}. Apart from eventually predicting promising new materials, the data produced by such a combinatorial \emph{high-throughput screening} (HTS) can be used to detect relationships between the composition, structure, and property of interest, which help to understand the underlying mechanisms better, and therefore may open up promising directions for optimization \cite{Hautier2011, Xiao2015}.

Present state-of-the-art battery technology could be improved significantly by replacing the liquid electrolyte by a highly ion-conducting and electrochemically stable solid-state electrolyte (SSE). Batteries consisting of SSEs could be stacked more densely, and safety concerns such as temperature stability, flammability and leakage of toxic liquids would be overcome \cite{Masquelier2011, Takada2013}. However, currently known SSE materials don't yet reach the ionic conductivity of liquid electrolytes, and therefore they lead to lower power densities and slower charge/discharge rates \cite{Li2015}. Fast and reversible migration of the charge carrying ions is not only relevant for SSEs, but also for solid intercalation electrode materials when the battery is operating in the ''rocking chair'' mode, where the ions are repeatedly inserted into and extracted out of the cathodes or anodes \cite{Wu2017, Goriparti2014}.

The material class with the crystal structure of NaZr$_2$(PO$_4$)$_3$ (\emph{NZP}), also named \emph{NASICON} after the \emph{$\underline{Na}$ $\underline{S}$uper $\underline{I}$onic $\underline{Con}$ductor} Na$_{1+x}$Zr$_2$Si$_x$P$_{3-x}$O$_{12}$ ($0\le x\le 3$) \cite{Hong1976}, has been widely examined due to its structural framework exhibiting a 3-dimensional network of paths for ionic migration \cite{Anantharamulu2011, Guin2015}. Depending on their electronic conductivity, NZP compounds were studied as intercalation cathode materials \cite{Aatiq2002, Aragon2015, Mason2014} or as solid-state electrolytes \cite{Feng2013, Bucharsky2015, Arbi2015}, both for Na as well as for Li ions. The NZP family is an ideal candidate system for a systematic computational screening analysis with respect to structural stability and ionic migration properties. On the one hand, the rhombohedral NZP structure with space group $R\bar{3}c$ is known to be stable under a variety of different occupations of the cationic sublattices \cite{Hagman1968, Delmas1981, Aono1990, Petkov2003}; on the other hand, these occupations strongly influence the ionic conductivity \cite{Tranqui1981, Winand1990, Alamo1993, Kumar2016, Xu2017}. The latter is assumed to originate from differently sized bottlenecks along the migration paths of the mobile ionic species (e.g. Li), influenced by differently sized oxygen polyhedra around larger or smaller cations \cite{Kohler1983, Kohler1985, Losilla1998, Martinez-Juarez1998}. For the NZP-type compound LiTi$_2$(PO$_4$)$_3$ (LTP), it was shown by atomistic calculations that partial substitution of Ti by iso- or aliovalent cations changes the activation energy of vacancy-mediated Li migration due to different volumes of the oxygen octahedra around the Li atoms \cite{Lang2015}.

In this work, a variety of NZP-type compounds of the form LiX$_2$(LO$_4$)$_3$ containing a systematic set of nine different metal ions on the \emph{X} site, and either P or Mn on the \emph{L} site is examined with respect to the conductivity of Li ions. The energy barriers for vacancy-mediated migration processes are calculated by means of \emph{density functional theory} and \emph{nudged elastic band} methods \cite{Henkelman2000}. The prevalent expectation of the bottleneck in the migration path of NZP-type compounds crystallizing in the $R\bar{3}c$ structure is critically addressed. Relationships between the constituent elements, the local arrangement of ions, and the activation energies are discussed. The paper is organized as follows: in Sec.\,\ref{tech}, the details of the NZP structure are described, the computational methods are introduced, and the choice of elements and compounds for this study is motivated. A discussion of the coordination of the so-called M2 site, for which the observations reported in the literature are ambiguous, is given in Sec.\,\ref{m2}. Minimum energy paths for vacancy-mediated migration of Li ions are presented in Sec.\,\ref{energ} and discussed with regard to the bottleneck concept by comparing experimental and computational results. In Sec.\,\ref{act}, the dependence of activation energies of the migration process on the volumes of oxygen octahedra around the initial and final Li positions is investigated and a simple model is presented to explain the obtained correlations. Relationships between the composition, i.e. the elemental occupation of the cationic positions X and L in the NZP structure, and the local arrangement of oxygen cages are given in Sec.\,\ref{compo}. Our conclusions are given in Sec.\,\ref{conc}.

\section{Technical Details}\label{tech}

\subsection{The NZP Structure}\label{struct}

\begin{figure}
\centering
\includegraphics[width=12cm]{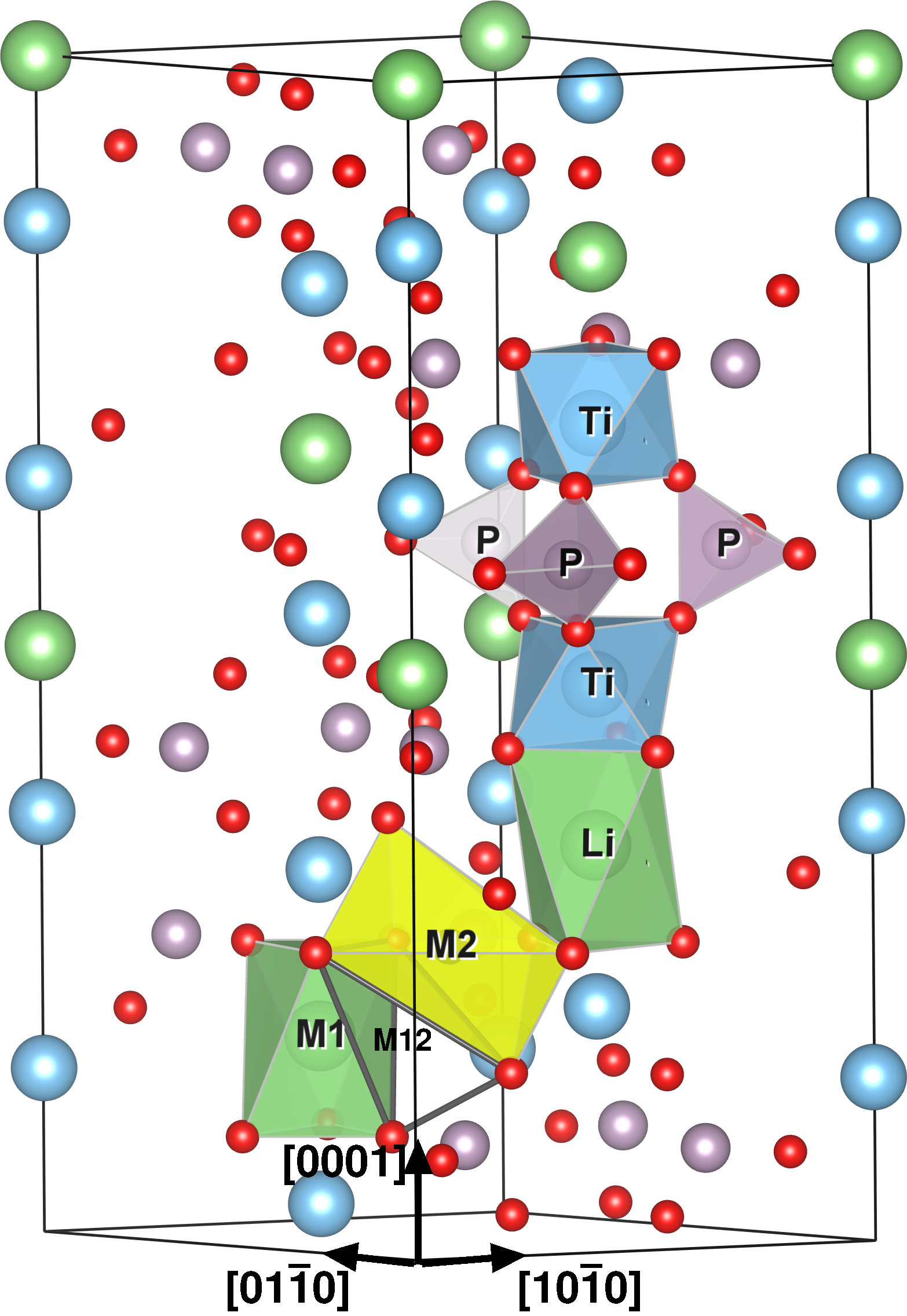}
\caption{Hexagonal cell of the rhombohedral NZP structure with the space group $R\bar{3}c$ for the example of LiTi$_2$(PO$_4$)$_3$ (containing 108 atoms corresponding to six formula units). Oxygen octahedra are surrounding Li on the M1 positions (green). In addition, oxygen octahedra around Ti on the X positions (blue), and oxygen tetrahedra around P on the L positions (purple) are shown, which build up a so-called lantern unit. The six-fold coordinated M2 position is located in the geometric center of the yellow octahedron. Solid gray lines between oxygen atoms highlight the tetrahedral environment between M1 and M2 octahedra, known as M12 position.}
\label{fig_01}
\end{figure}

NaZr$_2$(PO$_4$)$_3$, the prototype of the material class NZP with the general chemical formula M$_n$X$_2$(LO$_4$)$_3$ ($n=1,2,3,4$), crystallizes in the trigonal crystal system with space group $R\bar{3}c$. The M sites are occupied by monovalent cations such as Na$^{+}$, Li$^{+}$, or Cu$^{+}$ representing the mobile ionic species. Higher valent cations on the X positions (e.g. Al$^{3+}$, Ti$^{4+}$) and on the L positions (e.g. Si$^{4+}$, P$^{5+}$) build up the structural framework, the latter forming anionic groups (e.g. [SiO$_{4}$]$^{4-}$, [PO$_{4}$]$^{3-}$) with oxygen. The crystal structure belongs to the rhombohedral lattice system, but is generally represented with hexagonal lattice vectors. Fig.\,\ref{fig_01} shows the hexagonal cell containing six formula units of the NZP compound LiTi$_2$(PO$_4$)$_3$, i.e. 108 atoms. The structure is built up by so-called \emph{lanterns} consisting of two oxygen octahedra around the X (here Ti) atoms (Wyckoff position $12c$), and three oxygen tetrahedra around the L (here P) atoms ($18e$). There are two oxygen atoms in the asymmetric unit on Wyckoff positions $36f$ which form one substructure of triangular prisms building the cores of the lantern units, and a second substructure of octahedra being located above and below the lantern units in [0001] direction. The centers of these octahedra ($6b$) are called M1 sites, and they are occupied by M (here Li) atoms. Since the volume of an oxygen prism is only about 60\% of the M1 volume, the centers of the prisms ($6a$) remain unoccupied. In case of an occupation of lower valent cations on X or L sites, the amount of M atoms is increased in order to fulfill the condition of charge neutrality. These additional M atoms are occupying so-called M2 positions ($18e$) in between two neighboring M1 sites.

 The oxygen coordination of M2 is ambiguously reported in the literature; this will be discussed in detail in section \ref{m2}. Another interstitial position for M atoms is the M12 site (also labeled M3 in some references \cite{Arbi2015, Monchak2016}) on the general Wyckoff position $36f$, which is located in between neighboring M1 and M2 sites, i.e. within the distorted tetrahedra confined by the faces of the M1 and M2 octahedra (see Fig.\,\ref{fig_01}). Each M1 site is therefore surrounded by six M12 sites, which are occupied only pairwise by a ''dumbbell'' of M atoms due to symmetry reasons. In this case, the central M1 site is empty. Whether M1, M2, M12 or a mixture of these sites are occupied in an NZP compound depends on the type and valence of the cations. For example, in Li$_{1+x}$Al$_x$Ti$_{2-x}$(PO$_4$)$_3$ \cite{Arbi2015}, Na$_3$Sc$_2$(AsO$_4$)$_3$ \cite{Harrison2001}, and CuZr$_2$(PO$_4$)$_3$ \cite{Bussereau1992}, which are all crystallizing in the $R\bar{3}c$ structure, the M site atoms Li, Na and Cu are found on M1/M12, on M1/M2, and only on M12 sites, respectively.

In this work, the migration of a Li ion moving from a M1 site (Li$_{\mbox{\scriptsize M1}}$) via a vacant M2 site (Vac$_{\mbox{\scriptsize M2}}$) to a neighboring vacant M1 site (Vac$_{\mbox{\scriptsize M1}}$) is considered. The initial state ($\mbox{Li}_{\mbox{\scriptsize M1}}-\mbox{Vac}_{\mbox{\scriptsize M2}}-\mbox{Vac}_{\mbox{\scriptsize M1}}$) and transition state (Vac$_{\mbox{\scriptsize M1}}-\mbox{Li}_{\mbox{\scriptsize M2}}-\mbox{Vac}_{\mbox{\scriptsize M1}}$) configurations of this process will be labeled by $\mathcal{I}$ and $\mathcal{T}$, respectively, in the following. The energy difference, which is the activation energy for the migration process, is then defined as $E_{\mbox{\scriptsize a}} := E^{\mathcal{T}} - E^{\mathcal{I}}$.

\subsection{Choice of Compounds}\label{cmpds}

In a previous study, a computational high-throughput screening of 450 NZP compounds of the form LiX$_2$(LO$_4$)$_3$ with a variety of different elements on the X and L positions was conducted by the authors of the present work with the intention to identify new thermally stable NZP materials \cite{Mutter2017}. In detail, a systematic multi-step approach of combined molecular dynamics simulations for temperatures up to 800 K, the application of a stability criterion from the bond valence theory \cite{Brown1992}, and structural relaxation by DFT, where at each step all structures deviating from the rhombohedral NZP-type structure were sorted out, finally lead to a collection of compounds with a variety of metallic elements on the X position, and mostly P, Cr and Mn on the L position. Within this approach, all these compounds therefore are stable in the NZP structure even up to higher temperatures, at least in a local energy minimum.

For the present study, compounds with the 3d transition metal elements Ti, V, Fe, the 4d elements Zr, Nb, Ru, and the 5d elements Hf, Ta, and Os on the X position were taken into account, allowing for a systematic analysis across the periods 4, 5, and 6, and the groups 4, 5, and 8 of the periodic table. In order to analyze the influence of different L-site atoms, P and Mn were considered, leading to 18 compounds altogether. The choice of the X elements can further be justified since they are all found in stable materials (the XO$_2$ oxides) in the charge state +4 in an octahedral oxygen coordination, which is the same structural motive as in LiX$^{+4}_2$(L$^{+5}$O$_4$)$_3$. While tetrahedrally coordinated P$^{5+}$ configurations are widely present, compounds containing Mn$^{5+}$ cations in a tetrahedral oxygen environment appear unusual, but such an environment is present as well, e.g. in the stable chromophoric materials Ba$_3$Mn$_2$O$_8$ \cite{Weller1999, Laha2011} and Ba$_6$Na$_2$(Nb,Ta)Mn$_2$O$_{17}$ \cite{Han2017}.

\subsection{Methods of Computation}\label{meth}

Density functional theory (DFT) as implemented in the \emph{PWscf} code of the open-source software Quantum Espresso \cite{QE-2009} was applied to calculate the total energies of the NZP-type crystals, to relax the structures by means of a BFGS algorithm \cite{Shanno1970}, and to determine activation energies of the Li migration process. Corresponding energy barriers were obtained by applying the nudged elastic band method (NEB) \cite{Henkelman2000}. The general gradient approximation of Perdew \emph{et al.} (PBE) \cite{Perdew1997} was used for exchange-correlation. The Brillouin zone integrations in reciprocal space were carried out on a $3\times 3\times 1$ Monkhorst-Pack $k$-point mesh \cite{MonkhorstPack} with a Gaussian smearing of the partial orbital occupancies with a width of 0.01 Ry. The interaction of valence electrons with core electrons and atomic nuclei were taken into account by ultrasoft pseudo potentials provided by the GBRV library \cite{Vanderbilt1990, GBRV}. Wavefunctions of the valence electrons were expanded in a basis set of plane waves up to an energy cutoff of 35 Ry. To model the NZP compounds, a hexagonal cell consisting of 108 atoms (six NZP formula units) in the defect-free crystal structure was used as depicted in Fig.\,\ref{fig_01}. Each defect-free system was initially relaxed, i.e. the atomic positions were optimized until the minimal force component acting on an atom was lower than 0.002 Ry/$\mathring{\mbox{A}}$, and the volume was relaxed by minimizing the total energy and the internal stress. From these optimized cells of the perfect crystals, systems containing Li vacancies or interstitials were generated by extracting or inserting the atoms at the positions of interest, followed, unless otherwise stated, by only ionic relaxation at fixed cell volume with the same convergence criteria. Note that this workflow follows the same procedure as described in Ref.\,\cite{Lang2015}.

\section{Results and Discussion}\label{res}

\subsection{Coordination of the M2 Site}\label{m2}

\begin{figure}
\centering
\includegraphics[width=\linewidth]{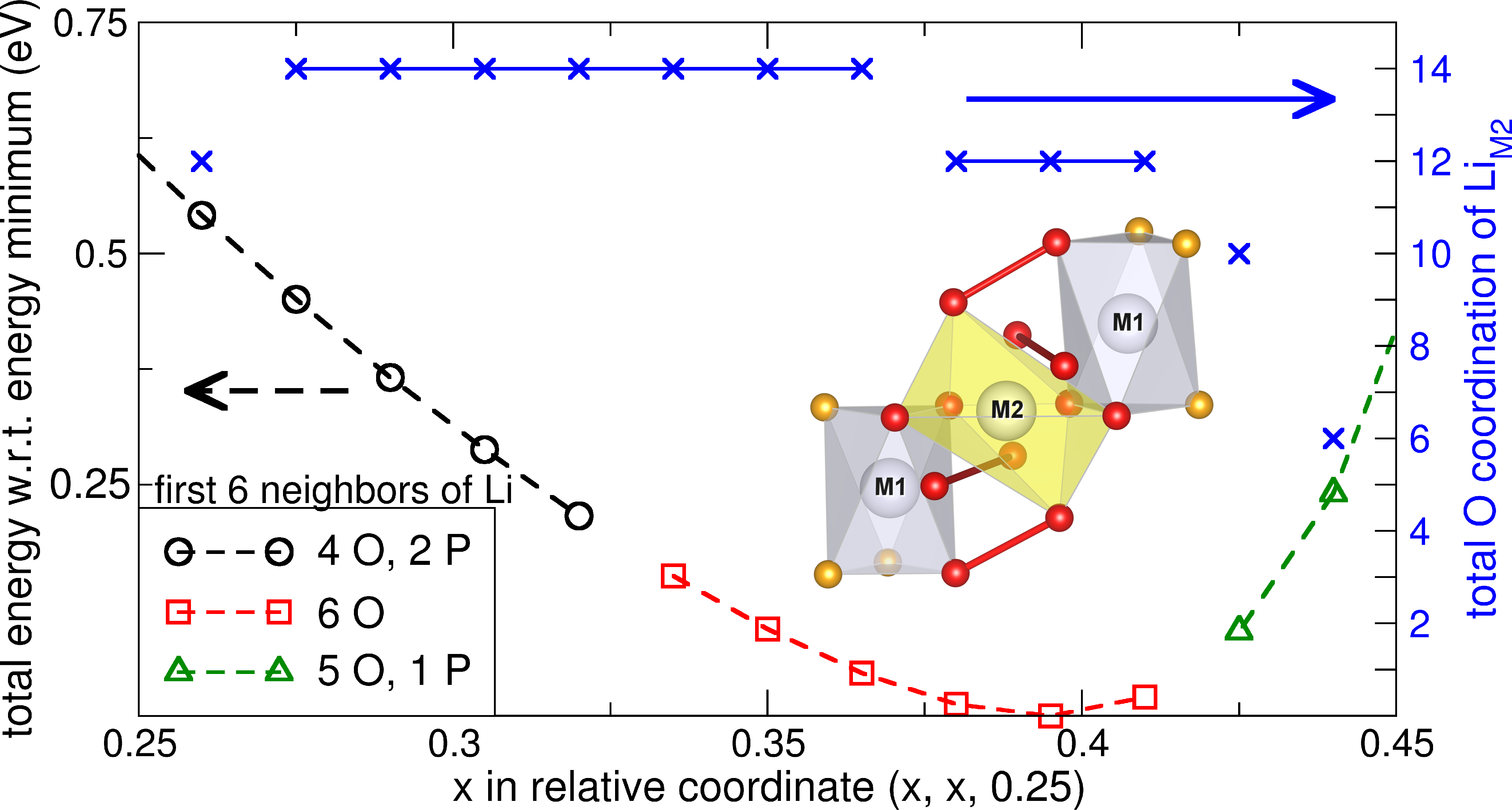}
\caption{Change in total energy (open symbols, left y-axis) and total oxygen coordination around the M2 site (crosses, right y-axis) of the compound LiTi$_2$(PO$_4$)$_3$, when a Li atom occupies the symmetry position $(x,x,0.25)$ representing a M2 site (Wyckoff position 18e), for varying values of $x$. Different open symbols distinguish the different possibilities of the closest six neighbors around the M2 site (see legend). No ionic relaxation was performed to obtain the energies. The inset shows the oxygen atoms surrounding the M1 sites (orange and red spheres) and the M2 site (red spheres).}
\label{fig_02}
\end{figure}

The M2 site in NZP-type compounds crystallizing in the space group $R\bar{3}c$ has been reported to be coordinated by 14 \cite{Alamo1993}, 10 \cite{Lang2015}, 8 \cite{Tranqui1981, Masquelier2000}, 6 \cite{Harrison2001}, or 4 \cite{Aragon2015} oxygen atoms. Since the M2 coordination is important for the discussion of the activation energies, it is analyzed here in detail. In general, the M2 site has the Wyckoff position $18e$ being represented for example by the relative coordinate $(x,x,0.25)$ in the asymmetric unit. The value of $x$ depends on the elements on the M, X, and L positions. It generally ranges from 0.3 to 0.4 (see e.g. Refs.\,\cite{Harrison2001, Masquelier2000, Lucazeau1986}). Exemplarily for LiTi$_2$(PO$_4$)$_3$ with Li occupying the M1 sites, an additional Li atom was placed on a M2 site with a varying value of $x$ between 0.25 and 0.45. The analysis of the neighbor environment yielded that for $x$ values between 0.33 and 0.41, which covers almost the full range of experimentally measured data, the 6 closest neighboring atoms are oxygen atoms forming a distorted octahedron (see Fig.\,\ref{fig_02}). Calculated total energy differences (without structural relaxation) further indicated that such a configuration is energetically favored. The \emph{total} oxygen coordination, as defined by the maximum sized polyhedron which contains no other atom inside, takes values between 14 and 12 in the range of experimental $x$ values. The 12-cornered polyhedron is indicated by the red colored oxygen atoms in the inset of Fig.\,\ref{fig_02}. Since some of these oxygen atoms are at rather long distances, up to about 4 $\mathring{\mbox{A}}$ from the M2 site, which is considerably longer than some of the distances to atoms of the other types (here Li, Ti, and P), these oxygen atoms presumably weren't counted to the number of coordinating atoms in the papers referenced above stating lower coordinations.

\subsection{Energy Barriers for Li Migration}\label{energ}

M-site ions in NZP-type structures are known to migrate along the interconnected network of M1, M12, and M2 sites. In order to determine the energies corresponding to a Li ion moving from a M1 site via M2 to a neighboring vacant M1 site, the initial-state [($\mathcal{I}$), see section \ref{struct}] and transition-state ($\mathcal{T}$) configurations were set up and structurally relaxed for all the 18 compounds considered in this study. Since the configurations correspond to either a local energy minimum ($\mathcal{I}$) or to a saddle point ($\mathcal{T}$), these relaxations did not lead to major atomic rearrangements except for an increase of the octahedron volumes around the vacant M1 sites between 8\% and 36\% compared to the perfect crystals. The volumes of the octahedra around all of the other still occupied M1 sites changed only negligibly, which further indicates the stability of the NZP-type structures. For the $\mathcal{T}$-configurations, the Li atom was initially placed on the M2 site $(x,x,0.25)$ with $x=0.38$, and the $x$ changed to values between 0.40 to 0.42 during relaxation, all lying in the minimum region depicted in Fig.\,\ref{fig_02}. This shows that the discussion above about the M2 coordination in LiTi$_2$(PO$_4$)$_3$ applies to the other considered NZP compounds as well. With these obtained configurations as initial and final states, NEB minimum-energy-path calculations were performed with three intermediate images which were initially set up along a straight line between the two limiting configurations. Note that due to the mirror symmetry of the energy profiles along the paths Li$_{\mbox{\scriptsize M1}}\rightarrow\,$Li$_{\mbox{\scriptsize M2}}$ and Li$_{\mbox{\scriptsize M2}}\rightarrow\,$Li$_{\mbox{\scriptsize M1}}$, it is sufficient to treat only one of these two sections explicitly.

\begin{figure}
\centering
\includegraphics[width=\linewidth]{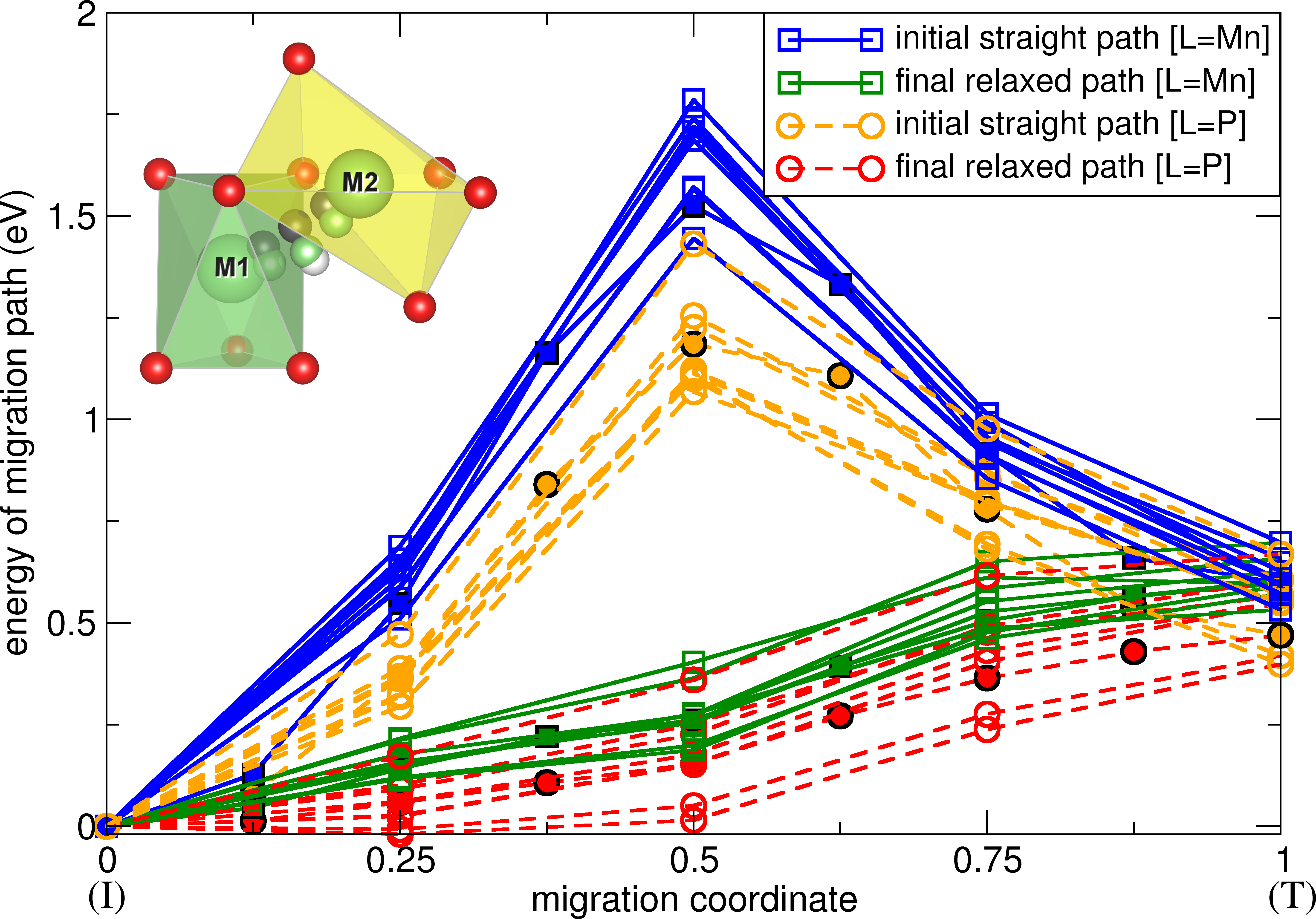}
\caption{Energy barriers for Li migrating between a M1 site [configuration ($\mathcal{I}$)] and the closest M2 site [configuration ($\mathcal{T}$)] for the compounds LiX$_2$(LO$_4$)$_3$ with X = Ti, V, Fe, Zr, Nb, Ru, Hf, Ta, Os and L = P, Mn. Results before and after the NEB calculation are shown. In addition to the migration paths obtained using NEB with 3 intermediate images, the paths for LiTi$_2$(PO$_4$)$_3$ and LiTi$_2$(MnO$_4$)$_3$, derived by applying 7 intermediate images, are depicted by black-circled symbols. The lines connecting the data points are given as a guide for the eye. The inset visualizes the oxygen octahedra around M1 and M2 positions, the initial straight path (small black spheres), and the final relaxed path (small green spheres). The small grey sphere represents the geometric center of the surrounding oxygen tetrahedron.}
\label{fig_03}
\end{figure}

Vacancy-mediated ionic migration of Na and Li ions in NZP-type compounds and the corresponding activation energies have been considered to be characterized by a bottleneck defined by the triangular faces of the oxygen coordinated M1 octahedra across which the ions have to move to reach the M2 sites \cite{Kohler1985, Martinez-Juarez1998, Losilla1998}. In this picture, one would expect at least one energy maximum along the M1$\rightarrow\,$M2 migration path, and eventually a second one when the moving atom crosses the face of the M2 coordinating octahedron. However, the evaluation of the minimum-energy paths before and after the NEB calculation, as shown in Fig.\,\ref{fig_03} for all the compounds, indicates, that there is only one distinct maximum corresponding to the central image of the initial path. This maximum considerably drops during the NEB calculation and ends up even below the energy of the transition-state configuration with Li on the M2 site. Since the energy curves are qualitatively similar except for variations in the barrier heights for all of the considered elemental substitutions, the behavior can be regarded as generally valid for NZP-type compounds with $R\bar{3}c$ symmetry. As can be seen in the inset of Fig.\,\ref{fig_03}, the origin of the energy maximum is the relatively short distance of a Li ion to two oxygen ions at the center of the initial path (i.e. half way between M1 and M2), as well as the asymmetric location of this atom in the tetrahedral oxygen environment. Relaxation cures this unfavored configuration by shifting the Li ion much closer to the geometric center of the tetrahedron of the oxygen ions. Since these positions are indeed the M12 sites which are known to be occupied in several NZP compounds, a low energy could be expected. However, they don't appear as energy minima along the paths, since in stable configurations, M12 positions are generally occupied pairwise, i.e. by a dumbbell of two atoms due to symmetry reasons \cite{Arbi2015, Bussereau1992}. In order to ensure that no energy maxima were overlooked due to the path discretization being too coarse, the NEB calculations for LiTi$_2$(PO$_4$)$_3$ and LiTi$_2$(MnO$_4$)$_3$ were repeated with 7 instead of 3 intermediate images. This however did not lead to any significant changes in the shapes of the minimum-energy-path curves.

In Fig.\,\ref{fig_03}, the minimum-energy-path curves of the compounds LiX$_2$(LO$_4$)$_3$ were not distinguished from each other with respect to different elements on the X positions. The effect of composition on activation energies will be discussed in detail in Sec.\,\ref{compo}. The curves corresponding to Mn on L positions show considerably higher energies than the curves for P on L positions. This is more pronounced for the unrelaxed paths, but still valid after the relaxation for most of the curves. The activation energy $E_{\mbox{\scriptsize a}}$ as defined in Sec.\,\ref{struct} for the vacancy-mediated migration of Li is obtained from the maximum value of the relaxed paths depicted in Fig.\,\ref{fig_03}. Since none of the relaxed paths exhibits any further maximum, the prevalent concept of a migration bottleneck between the M1 and M2 sites needs to be critically reassessed.

The bottleneck concept in NZP structures was first described in detail by Kohler and Schulz for the NASICON compounds Na$_{\mbox{\scriptsize 1+z}}$Zr$_{\mbox{\scriptsize 2-y}}$Si$_{\mbox{\scriptsize x}}$P$_{\mbox{\scriptsize 3-x}}$O$_{\mbox{\scriptsize 12}}$ \cite{Kohler1983, Kohler1985}. Using X-ray diffraction at elevated temperatures, where the compounds were in the rhombohedral NZP structure, probability density functions (PDFs) of Na ions within the lattice were obtained from measured anharmonic temperature coefficients of the Na site. Close to the center between two maxima corresponding to the M1 and M2 positions, a minimum of the PDFs was obtained and interpreted as the bottleneck for Na migration by linking the PDFs to an effective potential with a maximum at the bottleneck position.

\begin{figure}
\centering
\includegraphics[width=\linewidth]{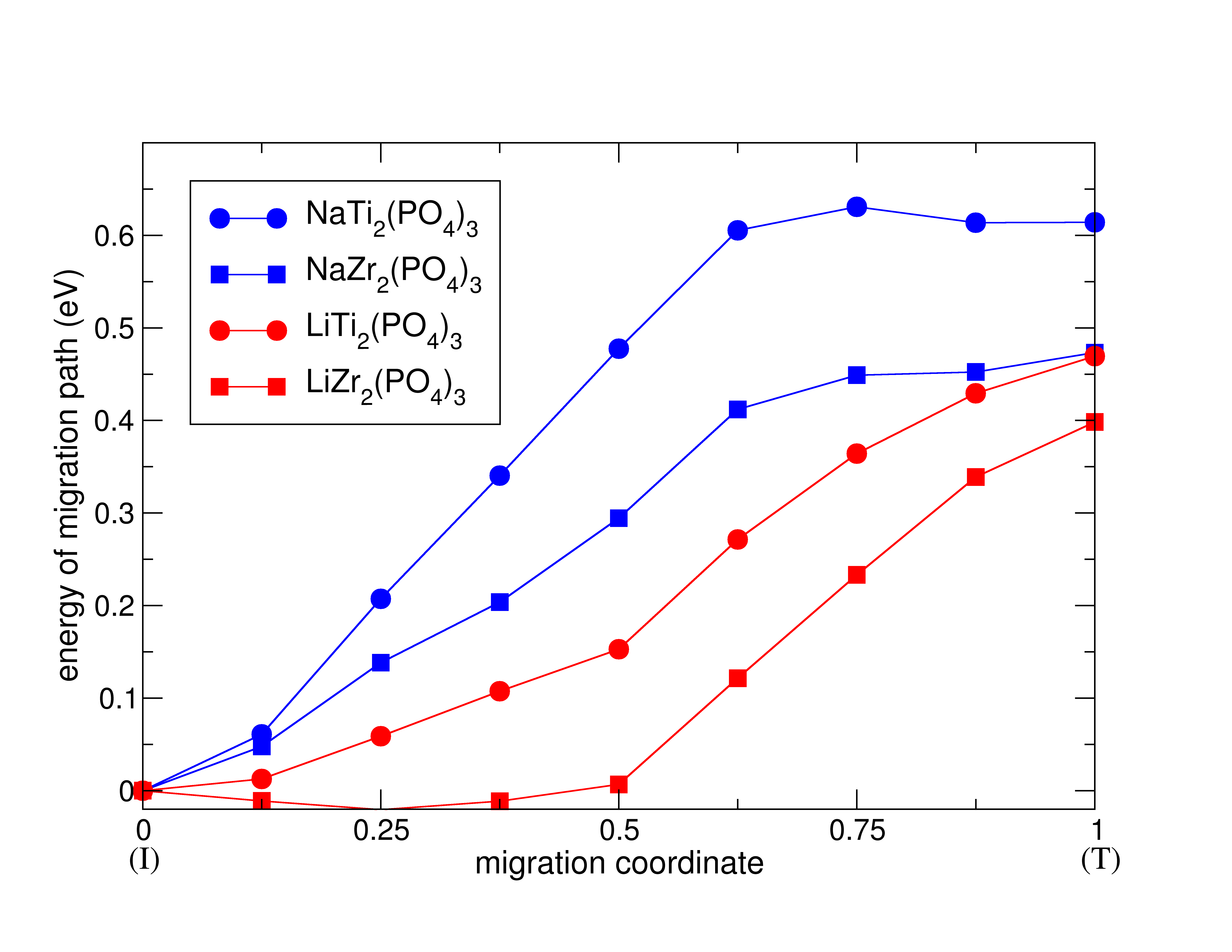}
\caption{Energies of the paths corresponding to Li [Na] migration between a M1 site [configuration ($\mathcal{I}$)] and a M2 site [configuration ($\mathcal{T}$)] for the compounds Li[Na]Ti$_2$(PO$_4$)$_3$ and Li[Na]Zr$_2$(PO$_4$)$_3$. The results are obtained by NEB relaxation with 7 intermediate images. The lines connecting the data points are given as a guide for the eye.}
\label{fig_04}
\end{figure}

As reported above, no such maximum appeared in the calculated NEB migration paths of the Li-containing NZP-type compounds. But since the radius of a Na$^+$ ion is by 34\,\% (for a 6-fold coordination) to 68\,\% (for a 4-fold coordination) larger than the radius of a Li$^+$ ion, one may assume distinct bottleneck-related maxima along the migration paths of Na in Na-containing NZP-type structures. Fig.\,\ref{fig_04} displays the paths for Na migration in NaTi$_2$(PO$_4$)$_3$ and NaZr$_2$(PO$_4$)$_3$ in comparison to those of Li migration in LiTi$_2$(PO$_4$)$_3$ and LiZr$_2$(PO$_4$)$_3$, all curves obtained in the same way as described in Sec.\,\ref{meth}. The flat region around the local minimum at the M2 position is more pronounced in the Na than in the Li compounds. But again, no pronounced maxima are found in the energy curves, which could be ascribed to structural bottlenecks along the paths.

The curve for LiZr$_2$(PO$_4$)$_3$ shows an unexpected minimum below the energy of the $\mathcal{I}$-configuration at a Li migration coordinate of 0.25. An additional relaxation confirmed the stability of this configuration, where the Li atom is located in the center between the M1 and the M12 position, corresponding to a shift of about 0.84\,$\mathring{\mbox{A}}$. LiZr$_2$(PO$_4$)$_3$ is reported to crystallize in a triclinic structure below 50$^{\circ}$C \cite{Catti1999}, so the $R\bar{3}c$ configuration has to be considered as a local minimum, which is prone to structural instabilities if lattice defects are present.

\begin{figure}
\centering
\includegraphics[width=\linewidth]{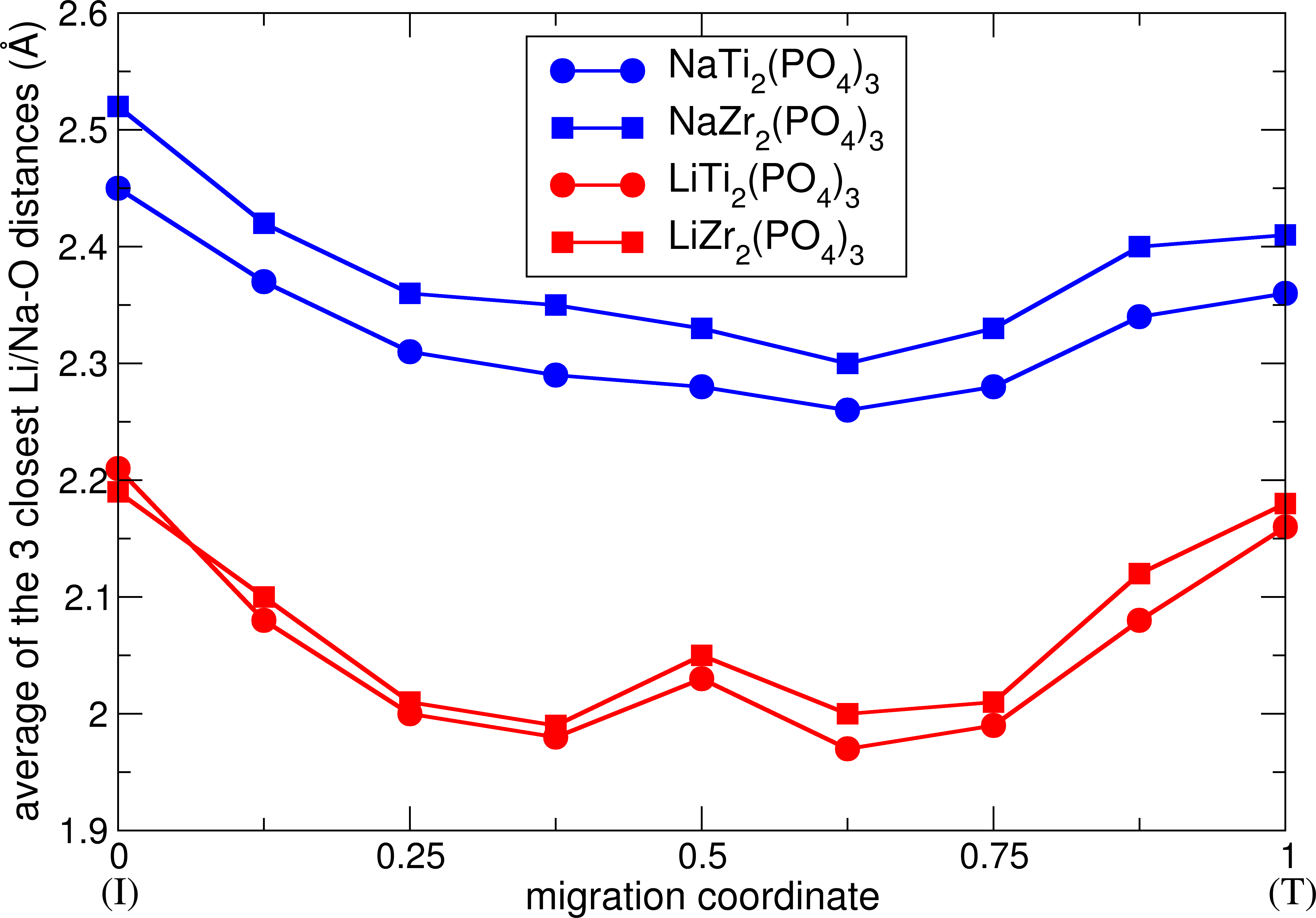}
\caption{Variation of the mean values of the distances between a Li[Na] atom and the three oxygen atoms next to it in the compounds Li[Na]Ti$_2$(PO$_4$)$_3$ and Li[Na]Zr$_2$(PO$_4$)$_3$ during the migration of Li[Na] between a M1 site (configuration $\mathcal{I}$) and a M2 site (configuration $\mathcal{T}$). The values are derived during the NEB relaxations corresponding to the energy paths as shown in Fig.\,\ref{fig_04}. The lines connecting the data points are given as a guide for the eye.}
\label{fig_05}
\end{figure}

 In analyzing the average distances of Li/Na to the three closest oxygen atoms (Fig.\,\ref{fig_05}) and setting them in relation to the energy barriers (Fig.\,\ref{fig_04}), no obvious correlation between the shape of the energy barriers and the atomic distances between Li or Na and O can be identified. Rather, the stable positions of Li or Na at the M1 and M2 positions can be linked to larger atomic distances as compared to lower atomic distances at the unstable intermediate path positions. Since at the intermediate positions, Na or Li ions are unlikely to be found, this could explain the above mentioned minima in the PDFs reported by Kohler and Schulz \cite{Kohler1983, Kohler1985}.

Results of DFT NEB calculations performed for various NZP-type compounds by Lang \emph{et al.} (\cite{Lang2015}), Shi \emph{et al.} (\cite{Shi2014}), and Lu \emph{et al.} (\cite{Lu2017}) do not show bottleneck-related energy maxima along Li or Na migration path either, but pronounced maxima at or close to the M2 position. Shi \emph{et al.} report smaller energy barrier heights for larger values of the average bond length between Li or Na and the \emph{eight} closest oxygen neighbors around M2, which can be related to a correlation of lowered activation energies with increasing volumes of the M2-coordination polyhedron. This will be discussed in Sec.\,\ref{act}. Not just the closest three oxygen atoms determine the energy of Li or Na atoms along the migration path, but the extended coordination environment has to be taken into account. This is also discussed by Lu \emph{et al.}, who considered in addition to the M1-M2 path some further migration directions for Li/Na in the rhombohedral NZP structure, leading to remarkably different (higher) activation energies while retaining very similar areas of the oxygen triangles across which Li or Na atoms have to migrate.

It is interesting to note that studies of NZP compounds using classical atomistic simulation methods such as a rigid-ion Coulomb-Buckingham pair potential \cite{Padmakumar2002} or bond-valence (BV) analysis \cite{Mazza2001, Rao2012} report energy maxima along the ionic transition paths. The migration barriers for Na between the M1 and M2 positions in Na$_{1+x}$Zr$_2$Si$_x$P$_{3-x}$O$_{12}$ for different values of $x$ determined by Mazza \cite{Mazza2001} applying BV sums show broad maxima, resembling quite well the behavior of the average distance between Na and the three closest oxygen atoms along the M1-M2-path shown in Fig.\,\ref{fig_05} for the Na compounds. A similar BV result for a Li migration path was obtained by Rao \emph{et al.} for Li$_{1.3}$Al$_x$Sc$_{0.3-x}$Ti$_{1.7}$(PO$_4$)$_3$ \cite{Rao2012}. BV sums can be formulated in terms of pairwise rigid-ion interactions \cite{Adams2011}, so the atoms closest to the migrating Li or Na atom contribute the most to the interaction energy and the overall Li or Na coordination plays a minor role. The lack of a self-consistent electronic screening of the migrating ion in the rigid-ion pair potential models presumably is the reason for the differing results when using classical simulation methods instead of DFT.

\subsection{Activation Energies}\label{act}

\begin{figure}
\centering
\includegraphics[width=0.96\linewidth]{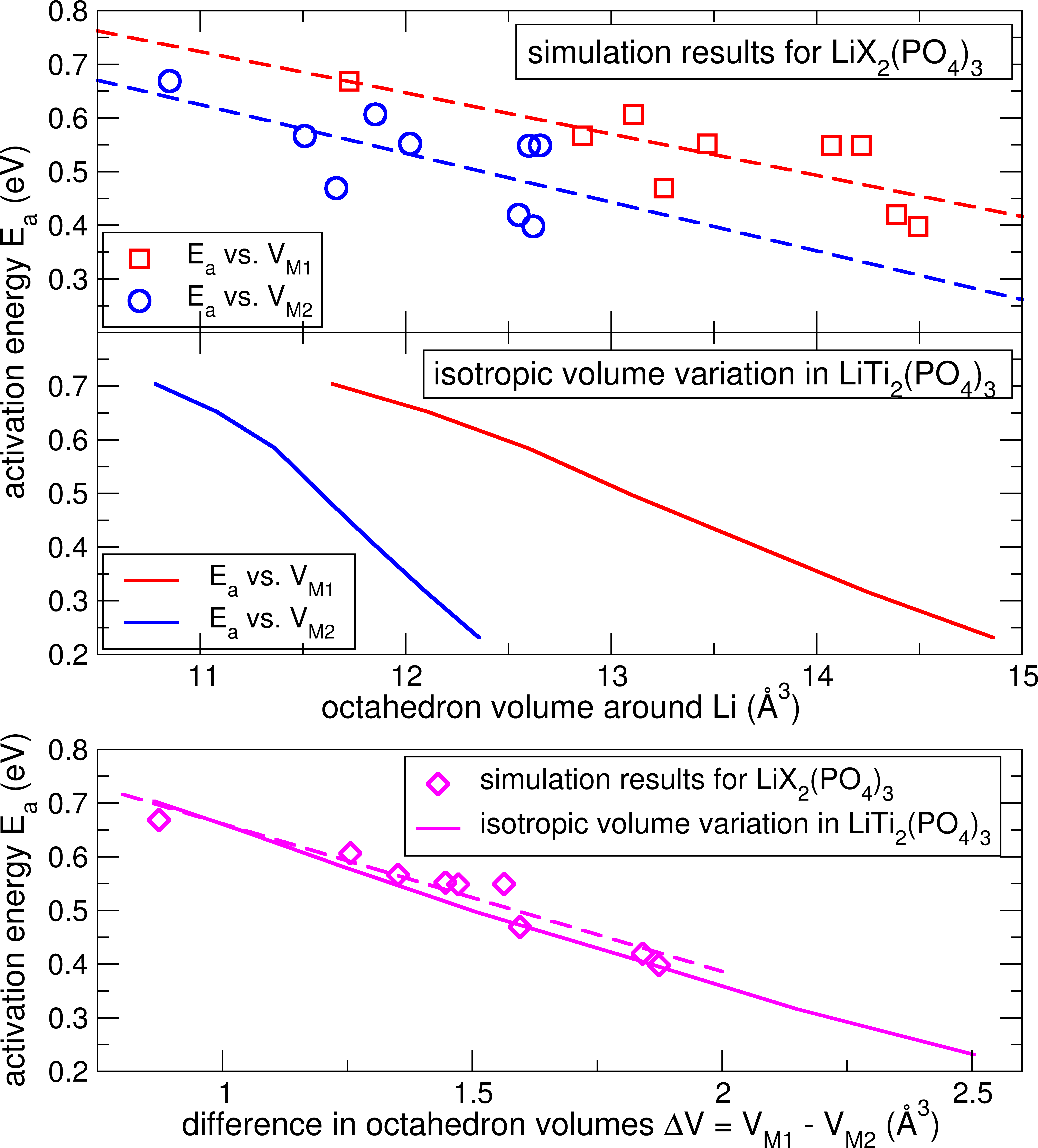}
\caption{$\underline{\mbox{Upper panel:}}$ Activation energies $E_{\mbox{\scriptsize a}}$ for vacancy-mediated Li migration between M1 and M2 sites in LXP compounds versus the oxygen octahedron volumes $V_{\mbox{\scriptsize M1}}$ around Li on M1 in configuration ($\mathcal{I}$), and versus $V_{\mbox{\scriptsize M2}}$ around Li on M2 in configuration ($\mathcal{T}$). The dashed lines are linear fits. $\underline{\mbox{Central panel:}}$ Dependence of $E_{\mbox{\scriptsize a}}$ on $V_{\mbox{\scriptsize M1}}$ and $V_{\mbox{\scriptsize M2}}$ for an isotropic volume variation in LiTi$_2$(PO$_4$)$_3$. $\underline{\mbox{Lower panel:}}$ Dependence of $E_{\mbox{\scriptsize a}}$ on $\Delta V = V_{\mbox{\scriptsize M1}} - V_{\mbox{\scriptsize M2}}$ for the LXP compounds (symbols; the dashed line is a linear fit), and for an isotropic volume variation of LiTi$_2$(PO$_4$)$_3$ (solid line).}
\label{fig_06}
\end{figure}

In the bottleneck concept, a clear correlation between the volumes $V_{\mbox{\scriptsize M1}}$ of oxygen octahedra around Li on a M1 position and the activation energies of ionic migration is expected. In the upper panel of Fig.\,\ref{fig_06}, the activation energies $E_{\mbox{\scriptsize a}}$ for the compounds LiX$_2$(PO$_4$)$_3$ (LXP) are shown versus the volumes $V_{\mbox{\scriptsize M1}}$ in configuration ($\mathcal{I}$) and versus the volumes $V_{\mbox{\scriptsize M2}}$ in configuration ($\mathcal{T}$). The regression lines indicate that there is indeed a trend of decreasing energy barriers with larger octahedron volumes, but the correlations are scarce, especially for the M2 volumes. Since the activation energies are calculated by the energy difference of the two terminating states at M1 and M2, a quantity taking into account characteristics of \emph{both} configurations ($\mathcal{I}$) and ($\mathcal{T}$) should rather correlate with $E_{\mbox{\scriptsize a}}$. Indeed, a clearly better correlation can be obtained between $E_{\mbox{\scriptsize a}}$ and the \emph{difference} in octahedron volumes $\Delta V = V_{\mbox{\scriptsize M1}} - V_{\mbox{\scriptsize M2}}$ (see Fig.\,\ref{fig_06}, lower panel).

In order to understand the structural influence of changing octahedron volumes on $E_{\mbox{\scriptsize a}}$ irrespective of the chemical influences of different atomic species, we studied the effect of an isotropic volume variation on the compound LiTi$_2$(PO$_4$)$_3$. The lattice constants $a$ and $c$ of the hexagonal cell were varied stepwise while keeping the $c/a$ ratio constant at its equilibrium value. At each volume step, configurations ($\mathcal{I}$) and ($\mathcal{T}$) were set up, and the internal coordinates were relaxed. It is found that the activation energies decrease with increasing octahedron volumes around Li on both M1 and M2 positions as depicted in Fig.\,\ref{fig_06} (central panel). The slopes of these curves differ from those of the regression lines through the values of all the LXP compounds (upper panel), but the dependence of $E_{\mbox{\scriptsize a}}$ on $\Delta V$ from the isotropic volume variation almost perfectly matches the line of $E_{\mbox{\scriptsize a}}$ versus $\Delta V$ through the LXP values (lower panel). This confirms the correlation between $E_{\mbox{\scriptsize a}}$ and $\Delta V$, which can be regarded as a general property of the underlying NZP structure, independent of the lattice parameters, with its special octahedral environments of Li in the initial state M1 and in the transition state M2.

\begin{figure}
\centering
\includegraphics[width=\linewidth]{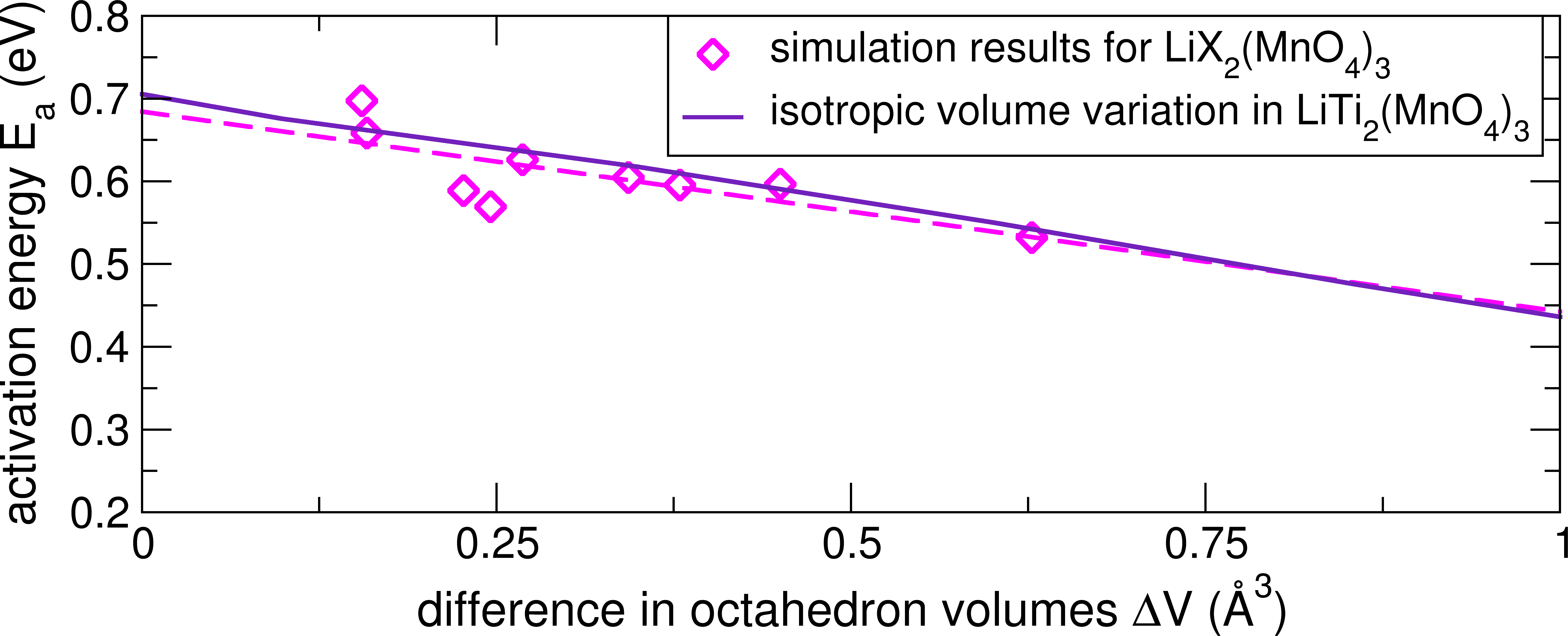}
\caption{Dependence of $E_{\mbox{\scriptsize a}}$ on $\Delta V$ for the LXM compounds (symbols; the dashed line is a linear fit), and for an isotropic volume variation of LiTi$_2$(MnO$_4$)$_3$ (solid line).}
\label{fig_07}
\end{figure}

The generality of this statement is underpinned by the fact, that a similar behavior was observed for the compounds containing Mn instead of P [LiX$_2$(MnO$_4$)$_3$ (LXM)] (see Fig.\,\ref{fig_07}), where the line through the $E_{\mbox{\scriptsize a}}$ values also matches the line resulting from an isotropic volume variation, in this case obtained for LiTi$_2$(MnO$_4$)$_3$, when correlated with the octahedron volume differences $\Delta V$. The slopes of the regression lines for L=P ($-0.27$ eV/$\mathring{\mbox{A}}^3$) and L=Mn compounds ($-0.24$ eV/$\mathring{\mbox{A}}^3$) differ by only 0.03 eV/$\mathring{\mbox{A}}^3$. This indicates that there is only a little range for the slope values of these curves, only defined by the underlying NZP crystal structure and the type of the migrating atom. However, the line for L=P and L=Mn have different y-axis intercepts, i.e. LXP and LXM compounds cannot be represented by a single line of $E_{\mbox{\scriptsize a}}$ versus $\Delta V$. This is because of the larger ionic radius of Mn$^{+5}$ compared to P$^{+5}$, leading to considerably lower LiO$_6$ volume differences between M1 and M2 octahedra for LXM than for LXP compounds. This will be discussed in more detail in Sec.\,\ref{compo}. The strongest deviations from the regression lines are obtained for LiTa$_2$(PO$_4$)$_3$ (42 meV), LiRu$_2$(MnO$_4$)$_3$ (54 meV), LiTa$_2$(MnO$_4$)$_3$ (40 meV), and LiNb$_2$(MnO$_4$)$_3$ (54 meV). As stated in Sec.\,\ref{cmpds}, the DFT approach predicts all the compounds considered in this study to be stable in the NZP-type structure in at least a local energy minimum. But if such an energy minimum is shallow, lattice imperfections such as Li vacancies can cause slight distortions of the LiO$_6$-octahedra in the vicinity. This may be a reason for the deviations of the points $E_{\mbox{\scriptsize a}}\left(\Delta V\right)$ found for some of the compounds of up to about 50 meV from the regression lines, which corresponds to a deviation of the volume difference of about 0.25 $\mathring{\mbox{A}}^3$.

\begin{figure}
\centering
\includegraphics[width=\linewidth]{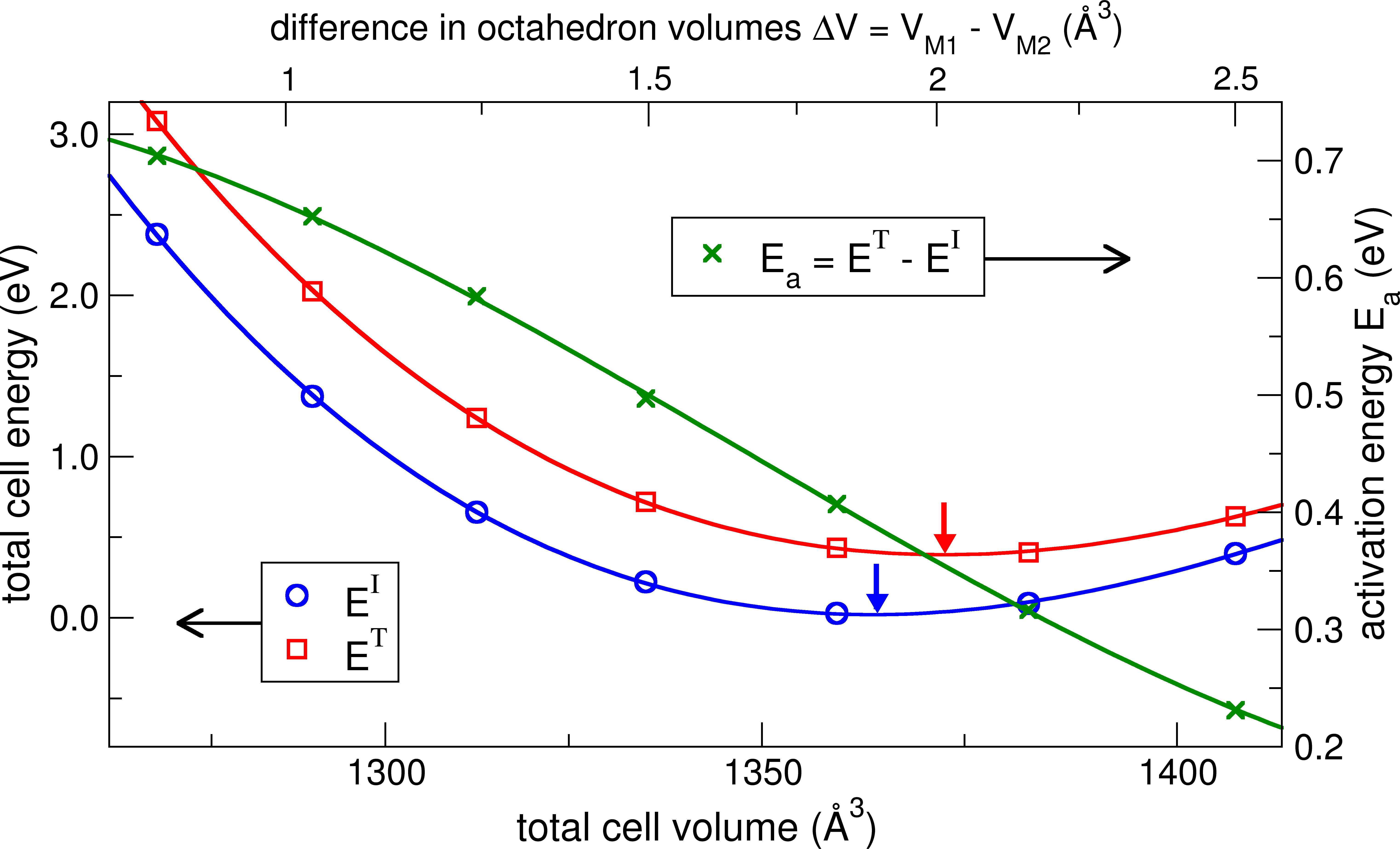}
\caption{Total energies (left y-axis) of the initial state configurations ($E^{\mathcal{I}}$, blue circles) and of the transition state configurations ($E^{\mathcal{T}}$, red squares) as a function of the total volume of the unit cell, as derived by an isotropic variation of the lattice parameters of a LiTi$_2$(PO$_4$)$_3$ crystal [see also Fig.\,\ref{fig_06} (central panel) and description in the text]. The variation of the activation energy $E_{\mbox{\scriptsize a}}$ (green crosses, right y-axis) is also given with respect to the cell volume. The difference $\Delta V = V_{\mbox{\scriptsize M1}}-V_{\mbox{\scriptsize M2}}$ in the oxygen octahedron volumes around Li on M1 and on M2 in configurations ($\mathcal{I}$) and ($\mathcal{T}$), respectively, depends on the total cell volume and is given on the upper x-axis. The lines are polynomial fits through the data points. Total energies are given with respect to the minimum of the $E^{\mathcal{I}}$ curve (equilibrium). The minima of the $E^{\mathcal{I}}$ and $E^{\mathcal{T}}$ curves are highlighted by arrows.}
\label{fig_08}
\end{figure}

In order to understand the decreasing $E_{\mbox{\scriptsize a}}$ with increasing $\Delta V$ when the volume is isotropically varied, we consider the dependence of the total energies $E^{\mathcal{I}}$ and $E^{\mathcal{T}}$, which define the activation energy, on a change of the total cell volume $V_{\mbox{\scriptsize c}}$ [see Fig.\,\ref{fig_08} for LiTi$_2$(PO$_4$)$_3$]. Both curves follow an equation of state, which can be approximated by a parabolic function around their respective minima. When Li sits on the M1 site, i.e., on its energetically most stable position in the center of a regular octahedron, the bonds to the surrounding oxygen atoms are stronger than for Li on the less favorable M2 site within a more distorted octahedron. Therefore, the curvature of the $E^{\mathcal{I}}\left(V_{\mbox{\scriptsize c}}\right)$ parabola is larger than that of the $E^{\mathcal{T}}\left(V_{\mbox{\scriptsize c}}\right)$ parabola, which corresponds to a higher bulk modulus of the initial compared to the transition state. Furthermore, the less strongly bound Li on M2 together with the expansion of the two adjacent empty M1 octahedra results in a larger equilibrium cell volume in the transition state. In combination, these two effects lead to a decreasing energy difference with increasing cell volume in the region around the equilibrium. This can already be deduced simply by considering the difference function $\Delta g\left(x\right):=g_2\left(x\right)-g_1\left(x\right)$ between two shifted parabolae $g_1\left(x\right)=x^2$ and $g_2\left(x\right)=a\left(x-\delta\right)^2$ $\left(\delta>0\right)$ with $g_2$ having a lower (positive) curvature than $g_1$, i.e. $0<a<1$. Under these conditions, $\Delta g\left(x\right)$ has a maximum in the negative $x$-range, and therefore decreases monotonically in the region around the minima of $g_1$ and $g_2$ at $x=0$ and $x=\delta$, respectively. Furthermore, from the crystal structure (Fig.\,\ref{fig_01}) it follows, that an isotropic change in the lattice parameters (or cell volume) has a larger influence on the size of the oxygen octahedra around M1 than on the size of the octahedra around M2. This can be derived formally, but also understood intuitively, since the octahedra around M1 are elongated in the [0001] direction of the hexagonal cell, while the octahedra around M2 are inclined relative to this axis. Altogether, increasing the cell volume results in a raising difference between the octahedron volumes around M1 and M2, \emph{and} in a lowering of the activation energy, which explains the observed good correlation of $E_{\mbox{\scriptsize a}}$ and $\Delta V$.

\subsection{Influence of Composition}\label{compo}

It is intuitively obvious that the migration capability of atoms along pathways within a defined crystal structure must somehow be related to the interatomic distances in the structure, being defined by its lattice parameters. These in turn are influenced by the ionic radii of the constituent atoms. Since NZP compounds remain structurally stable for a wide variety of different elements on the cationic sites, a dependence of ionic conductivity on composition and lattice dimensions can directly be studied by analyzing the effect of elemental substitutions on the activation energies. Such correlations were experimentally reported for Na \cite{Tranqui1981, Delmas1981, Winand1990, Losilla1998} as well as for Li \cite{Aono1990, Martinez-Juarez1998, Kumar2016, Xu2017} containing NZP compounds. For example, Losilla et al. \cite{Losilla1998} reported increasing lattice parameters and decreasing activation energies when In$^{+3}$ in Na$_3$In$_2$(PO$_4$)$_3$ was partially substituted by Ti$^{+4}$, Sn$^{+4}$, Hf$^{+4}$, and Zr$^{+4}$ (in this order), having ionic radii $r_{\mbox{\scriptsize ion}}$ of 0.61 $\mathring{\mbox{A}}$, 0.69 $\mathring{\mbox{A}}$, 0.71 $\mathring{\mbox{A}}$, and 0.72 $\mathring{\mbox{A}}$, respectively, in a six-fold coordination \cite{Shannon1976}. For LiM$_2$(PO$_4$)$_3$ an analogous behavior was observed by Juarez et al. for M = Ge ($r_{\mbox{\scriptsize ion}}=0.53\,\mathring{\mbox{A}}$), Ti, Sn, and Hf.

\begin{figure}
\centering
\includegraphics[width=\linewidth]{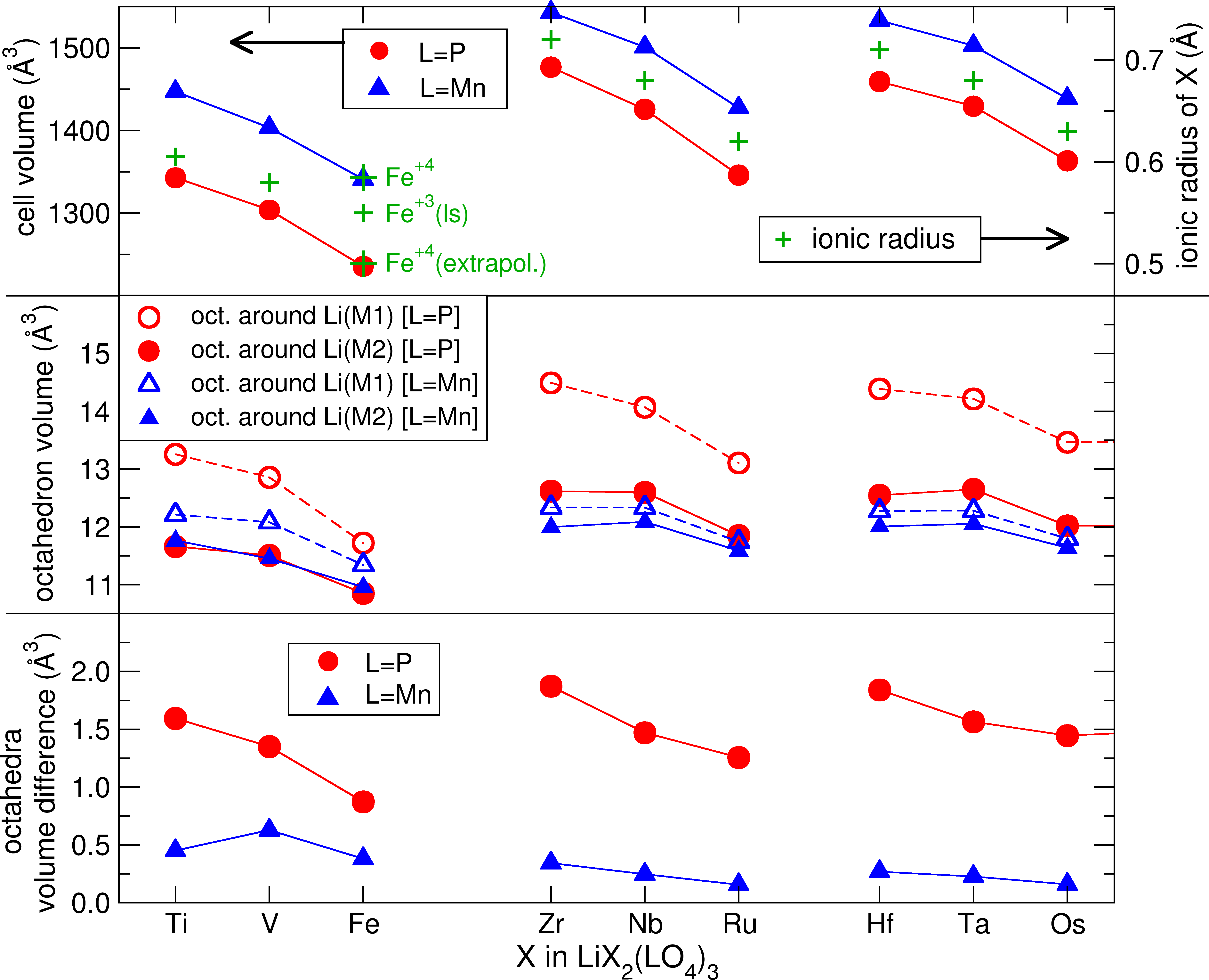}
\caption{The volumes of the perfect NZP-type crystal cells (i.e. without any vacancies on the M1 sites, and no Li on a M2 site) (upper panel, left y-axis), the volumes $V_{\mbox{\scriptsize M1}}$ [$V_{\mbox{\scriptsize M2}}$] of the oxygen octahedra around Li on M1 [M2] sites in configuration $\mathcal{I}$ [$\mathcal{T}$] (central panel), as well as the octahedron volume differences $\Delta V = V_{\mbox{\scriptsize M1}}-V_{\mbox{\scriptsize M2}}$ (lower panel) are given for all of the considered compounds LiX$_2$(LO$_4$)$_3$. Circles depict the results for L=P, triangles the results for L=Mn compounds. The arrangement of the data points for the different X elements on the common x-axis follows the periods and groups of the periodic table of elements from left to right and from top to bottom, respectively. Lines connecting the data points are given as a guide for the eye. In addition, in the upper panel, the ionic radii of the X$^{+4}$ ions in a 6-fold coordination are shown as given by Shannon \cite{Shannon1976} (green pluses, right y-axis). A description and discussion of the different values for Fe ions is given in the text.}
\label{fig_09}
\end{figure}

In our study, a tendency of decreasing activation energies with increasing volumes $V_{\mbox{\scriptsize M1}}$ and $V_{\mbox{\scriptsize M2}}$, and, considerably better correlated, with the octahedron volume differences was obtained (Fig.\,\ref{fig_06}). The dependence of these quantities on the composition, i.e. on the elements occupying the X and L positions in LiX$_2$(LO$_4$)$_3$, is shown in Fig.\,\ref{fig_09}. At first, in the upper panel of the graph, it is clearly visible that the total cell volume strongly correlates with the ionic radius of the atom on the X sites. The ionic radii were taken from the list given by Shannon \cite{Shannon1976} for ions in a charge state $+4$ and a 6-fold coordination. The ionic radius of Fe$^{+4}$ doesn't match the trend. However, in a six-fold oxygen coordination Fe is hardly found in the charge state $+4$ but rather in the state $+3$, as in Fe$_2$O$_3$ or Fe$_3$O$_4$. Furthermore, synthesized iron-containing NZP compounds are reported to contain more than one Li atom per formula unit, which is related to the charge state of Fe being $+3$ rather than $+4$ in these compounds \cite{Masquelier1998, Patoux2003}. The ionic radius of Fe$^{+3}$ in the low spin state (''ls'' in Fig.\,\ref{fig_09}) better fits to the correlation with the cell volume. On the other hand, when extrapolating both low-spin and high-spin Shannon radii for the charge states $+2$ and $+3$ to $+4$, one ends up at a value of 0.5 $\mathring{\mbox{A}}$. Following the trend of the cell volumes across the periodic table, the $r_{\mbox{\scriptsize ion}}$ of iron in both of the calculated compounds LiFe$_2$(PO$_4$)$_3$ and LiFe$_2$(MnO$_4$)$_3$ can be assumed to range between that value and the low spin value of Fe$^{+3}$ (0.55 $\mathring{\mbox{A}}$). With decreasing total cell volume, the volumes $V_{\mbox{\scriptsize M1}}$ and $V_{\mbox{\scriptsize M2}}$ decrease, too (see central panel of Fig.\,\ref{fig_09}). Since this affects $V_{\mbox{\scriptsize M1}}$ more strongly than $V_{\mbox{\scriptsize M2}}$, their difference $\Delta V$ drops for shrinking cell volumes, in agreement with the discussion for the isotropic volume change of a single compound (see Sec.\,\ref{act} and Fig.\,\ref{fig_08}). Only the compound LiTi$_2$(MnO$_4$)$_3$ exhibits a small deviation from this behavior.

The ionic radius of Mn$^{+5}$ (0.33 $\mathring{\mbox{A}}$) is about twice as large as that of P$^{+5}$ (0.16 $\mathring{\mbox{A}}$), which is reflected by larger total cell volumes. But the volumes $V_{\mbox{\scriptsize M1}}$ are considerably smaller in the LXM than in the LXP compounds. Compared to the X-O$_6$ configurations, where four electrons from the X atom contribute to the bonds, the Li$_{\mbox{\scriptsize M1}}$-O$_6$ octahedra with just one shared valence electron from the Li atom are less stiff under compression. An increasing size of L-O$_4$ tetrahedra will therefore push the stiff corner-connected X-O$_6$ octahedra away from each other while compressing the Li$_{\mbox{\scriptsize M1}}$-O$_6$ octahedra of the same lantern unit in [0001] direction (see Fig.\,\ref{fig_01}). Due to the face-connection of the Li$_{\mbox{\scriptsize M1}}$-O$_6$ to the stiff X-O$_6$ octahedra, the former won't be expanded in a direction perpendicular to [0001] by an increase of the ionic radius of the L-site atom. The whole lantern units will rather be pushed away from each other in [10$\bar{1}$0] and [01$\bar{1}$0] directions, leading to an increase of the corresponding lattice parameter ($a$). Since the expansion of the lattice parameter $c$ is almost compensated by the considerable shrinkage of the Li$_{\mbox{\scriptsize M1}}$-O$_6$ octahedra in [0001]-direction, the average $c/a$ ratio over the considered compounds drops from 2.41 for LXP to 2.34 for LXM. These cell deformations also influence the size of the Li$_{\mbox{\scriptsize M2}}$-O$_6$ octahedra. But due to counteracting effects in neighboring lantern units and the face-connection to two stiff X-O$_6$ octahedra, the volume changes are less pronounced, resulting in lower volume differences between Li$_{\mbox{\scriptsize M2}}$-O$_6$ and Li$_{\mbox{\scriptsize M1}}$-O$_6$ octahedra for larger ions on the L positions.

This observation for the influence of the L$^{+5}$-ions, and the systematic decrease of cell and Li-octahedron volumes as well as of volume differences for shrinking ionic radii of the X$^{+4}$ ions from left to right along periods of the periodic table of elements, together with the correlation of octahedron volume differences and vacancy migration energies of Li ions discussed above, constitute a useful composition-structure-property relationship for rhombohedrally crystallizing NZP-type compounds.

\section{Conclusions}\label{conc}

Using density functional theory and nudged elastic band calculations, the migration paths and activation energies for vacancy-mediated hopping of Li ions were determined in a systematic set of 18 NZP-type compounds LiX$_2$(LO$_4$)$_3$ with X = Ti, V, Fe, Zr, Nb, Ru, Hf, Ta, Os, and L = P, Mn. In the rhombohedral NZP structure, which from a previous study is known to be stable for all the considered compounds, vacancy-mediated Li migration can take place between M1 and M2 sites. The energy barriers show qualitatively similar behavior for the different cationic occupations with an energy minimum for Li occupying the M1 site and a maximum for Li at M2, but lacking any further intermediate maximum. The existence of such an intermediate maximum is prevalently assumed to originate from a close triangular oxygen arrangement (''bottleneck'') around Li when it crosses the face of the oxygen octahedron surrounding the M1 site. This concept was surveyed more deeply for selected Li and Na containing NZP compounds. It was found that along the M1-M2 migration path, there is no distinct energy minimum at the closest distances of the migrating Li or Na ion to the oxygen ions which could be related to a structural bottleneck with a corresponding maximum in the energy barrier.

Instead, the activation energy results obtained from the energy difference between Li on the M2 and M1 positions, and therefore the atomic environments of these two configurations need to be taken into account to correlate the local atomic environment of the moving ion with the activation energy. As a simple approach, it was shown that the difference between the volumes of oxygen octahedra surrounding M1 and M2 correlates linearly with the activation energy within 0.05 eV for the LiX$_2$(LO$_4$)$_3$ compounds, with similar slopes but different y intercepts for L=P and L=Mn. This linear correlation is reproduced when considering an isotropic volume variation in each one of the P and Mn compounds, which could be traced back to slight differences in the parabolic energy curves of a system in the initial (or final) state (Li on M1) and in the transition state (Li on M2). From left to right along a period in the periodic table, the ionic radius of the X$^{+4}$ cation decreases, and so do the NZP cell volumes, the volumes of M1 and M2 octahedra, and also mainly each of the octahedron volume differences, which links the composition to the structure in terms of local atomic environments of the migrating ion, and to the activation energy.

\begin{acknowledgments}
This work was funded by the German Research Foundation (DFG Grant no. El 155/26-1). The calculations were performed on the computing facility ForHLR I of the Steinbuch Centre for Computing (SCC) of the Karlsruhe Institute of Technology (KIT), funded by the Ministry of Science, Research, and Arts Baden-W\"{u}rttemberg, Germany, and by the DFG. Structure figures were prepared with VESTA \cite{vesta}.
\end{acknowledgments}


\end{document}